\documentclass[conference, anonymous]{IEEEtran}
\usepackage{cite}

\ifCLASSINFOpdf
   \usepackage[pdftex]{graphicx}
\else
   \usepackage[dvips]{graphicx}
\fi
\usepackage[cmex10]{amsmath}
\usepackage{algorithmic}

\usepackage{array}

\usepackage{mdwmath}
\usepackage{mdwtab}

\usepackage{eqparbox}

\usepackage[tight,footnotesize]{subfigure}

\usepackage{fixltx2e}

\usepackage{stfloats}

\usepackage{url}

\newtheorem{theorem}{Theorem}

\begin{document}
%
\title{Reconstruction of Aggregation Tree in spite of Faulty Nodes in Wireless Sensor Networks}


\author{\IEEEauthorblockN{Punit Sharma and Partha Sarathi Mandal}
\IEEEauthorblockA{
Indian Institute of Technology Guwahati\\ Guwahati - 781 039, India\\
Email: \{psm, s.punit\}@iitg.ernet.in}


}


%



\maketitle

\begin{abstract}

Recent advances in wireless sensor networks (WSNs) have led to many new promissing applications.
However data communication between nodes consumes a large portion of the total energy of WSNs.
Consequently efficient data aggregation technique can  help greatly to reduce power consumption.
Data aggregation has emerged as a basic approach in WSNs in order to reduce the number of
transmissions of sensor nodes over {\it aggregation tree} and hence minimizing the overall power
consumption in the network. If a sensor node fails during data aggregation then the aggregation
tree is disconnected. Hence the WSNs rely on in-network aggregation for efficiency but a single faulty node can
severely influence the outcome by contributing an arbitrary partial aggregate value.

In this paper we have presented a distributed algorithm that reconstruct
the aggregation tree from the initial aggregation tree excluding the
faulty sensor node. This is a synchronous model that is completed in
several rounds. Our proposed scheme can handle multiple number of faulty nodes as well.


\end{abstract}


\ifCLASSOPTIONpeerreview
\begin{center} \bfseries EDICS Category: 3-BBND \end{center}
\fi
%
\IEEEpeerreviewmaketitle

\section{Introduction}
A \textit{ wireless sensor networks} (WSNs) consist of a large number of spatially distributed autonomous
resource-constrained tiny sensor devices which are used to lead many new promising applications.
The applications for WSNs are varied, typically involving some kind of monitoring, tracking, or
controlling. Specific applications include: Habitat monitoring, Object tracking, Nuclear reactor
control, Fire detection, Traffic monitoring, etc.
However data communication between nodes consumes a large portion of the total energy of WSNs.
Consequently efficient data aggregation technique can  help greatly to reduce power consumption.
Data aggregation has emerged as a basic approach in WSNs in order to reduce the number of
transmissions of sensor nodes over {\it aggregation tree} and hence minimizing the overall power
consumption in the network.

Depending on the application, sensor nodes either report each and every measurement to a gateway
or sink, or they perform in-network aggregation: En route to the sink, nodes combine their own measurement
with the one of other nodes in proximity, e.g., their children on an {\it aggregation tree}
rooted at the sink and spanning over all sensors \cite{Lin}. A large fraction of WSNs requires only a periodic
collection of an aggregate value (e.g., count, sum, average, etc.), and can do so with low network overhead.
With in-network aggregation, rather than relaying individual measurements across multiple hops, each
node transmits a single packet, {\it summarizing} the data from an entire area of the WSNs.

Typically, there are three types of nodes in WSNs: leaf sensor nodes, aggregators,
and a querier (sink) \cite{Alzaid}. The aggregators collect data from a subset of the network,
aggregate the data using a suitable aggregation function and then transmit the aggregated
result to an upper aggregator or to the querier who generates the query. The querier is
entrusted with the task of processing the received sensor data and derives meaningful
information reflecting the events in the target field. It can be the base station or sometimes
an external user who has permission to interact with the network depending on the network architecture.
Data communications between sensors, aggregators and the queriers consume a large portion of the total
energy consumption of the WSNs.

Most of the works \cite{Alzaid,Chan,Gao,Haghani,Lin} in literature focused on secure aggregation in WSNs.
Secure aggregation means protecting data from attackers, where attackers intend to change the aggregation
value and mislead the sink (or base station) resulting in false aggregation. They considered {\it faulty} node
as an attacker or adversary that can compromise with sensor nodes by controlling their functionality and inducing
arbitrary deviations from the protocols. But in our proposed algorithm, a faulty node is considered as a physical fault.

A sensor node is called {\it faulty}, if it cannot be able to communicate with any other sensor node in the WSNs.
A sensor node may fail due to lack of battery power or some hardware failures.  We may consider node failure as a permanent failure.

If a sensor node fails during data aggregation then the aggregation tree is disconnected.
Hence the WSNs rely on in-network aggregation for efficiency but a single faulty node can
severely influence the outcome by contributing an arbitrary partial aggregate value to the sink.

In a typical application, a WSN is scattered in a region where it is meant to collect data through
its sensor nodes. We consider WSNs as a weighted communication graph, $G_c=(V,E)$ (say) where each sensor node
is a vertex belonging to a set $V$ and the communication link between two sensor nodes is defined as an edge belonging to a set $E$.
Here edge weight is the cartesian distance between two sensor nodes. One node can communicate with other nodes
directly if they are in its transmission range.

Using some distributed minimal spanning tree (MST) algorithm \cite{Gallager} it is possible to construct an initial aggregation
tree ($T_a$). If one node fails, then we assume, by some fault detection algorithm \cite{Chan}, that other
nodes which are directly connected with the faulty node can detect the fault and the aggregation tree is decomposed
into number of trees (disjoint-set of forest) with respect to the aggregation tree.

Our objective in the paper is as following: Given a weighted communication graph $G_c$ and corresponding aggregation tree $T_a$  with $n$ nodes, if one arbitrary node, $v_f$ (say) fails then how to reconstruct the aggregation tree with $n-1$ nodes in a
distributed way (excluding the faulty nodes), provided the reduced communication graph, $G^{'}_c=(V^{'},E^{'})$ is still connected after removal of the faulty node, $v_f$ where $V^{'}=V\setminus\{v_f\}$ and $E^{'}=E\setminus\{$ all edges are connected with $v_f\}$.

\subsection{Related Work:}
Chan {\it et al.} proposed a protocol \cite{Chan} where they considered corrupted node as
 a malicious aggregator node. According to their protocol the answer given by aggregator is a good
 approximation of the true value even when the aggregator and a fraction of the sensor nodes are corrupted.
In the paper\cite{Haghani} Haghani {\it et al.} considered adversary node as a misbehavior node that can
severely influence the outcome by contributing an arbitrary partial aggregate value. Their scheme relies on
costly operation to localize and exclude nodes that manipulate the aggreagtion when a fault is detected.
Gallager {\it et al.} \cite{Gallager} proposed a distributed algorithm (distributive implementation of Prim's algorithm)
constructing a MST of a connected graph in which the edge weights are unique. Their algorithm works on a
message passing model. It uses a bottom-up approach and the overall message complexity of the MST algorithm is $O(E + n{\rm lg}~ n$).
In the paper \cite{Gao} Gao and Zhu proposed a Dual-Head Cluster Based Secure Aggregation Scheme.
\subsection{Our results:}
The main contribution of this paper is a distributed algorithm for reconstruction of aggregation tree
in wireless sensor networks when an arbitrary sensor node fail during aggregation. To the best of our knowledge,
this is the first distributed protocol for reconstruction of aggregation tree which can handle multiple concurrent
permanent node failure. Unlike Gallager {\it et al.} \cite{Gallager} algorithm the edge weights of underline communication
graph may not be unique. We have proved that the reconstructed aggregation tree is again a MST. This is a synchronous model that completes in
several rounds. In terms of rounds the complexity of our algorithm are $O(1)$ in the best case, $O({\rm lg}~ n$) in the worst case.
The proposed algorithm can also handle multiple concurrent node failure.
\section{Reconstruction of Aggregation Tree}

Consider the connected WSN consisting of $n$ sensor nodes (vertices). Each sensor has its unique id, a
variable (initially zero), where edge weight is the communication cost between
two nodes. We assume that if one node fails the communication graph is still connected and by some fault detection algorithm neighbors of the
faulty node can detect the fault. We assume at a time there is only one faulty node in the WSN.
Our proposed algorithm is synchronous; i.e., its perform in several rounds. Due to failure of a node, the aggregation decompose in to disjoint set of forest (cluster, say). According to the algorithm each cluster will find the minimum outgoing edge (synchronously) and tries to merge with the
cluster on the other side of the edge. This is a distributed algorithm based on message passing.

\subsection{Notations}

Following notations are used throughout the paper for different type of message.
These message are required during execution of the algorithm.

\begin{itemize}
 \item $find\_msg$ (Find message): Fault detective node (cluster $root$, say) initiates  the message within the cluster to invoke the node(s) for finding $moe$.
 \item $report\_msg$ (Report message): Every leaf node in the cluster sends a $report\_msg$ with $moe$ information and own id to its parent after finding $moe$ from it, and every intermediate node sends $report\_msg$ to its parent after getting information about the $moe$ of its subtree including itself.
\item $test\_msg$ (Test message): A node issue a $test\_msg$ message through the $moe$ to know whether this edge is going to some other cluster.
\item  $accept\_msg$ (Accept message): A node generates a $accept\_msg$ message after receiving $test\_msg$ message if the $test\_msg$ message sender is belonging to different cluster.
\item $reject\_msg$ (Reject message): A node generates a $reject\_msg$ message after receiving $test\_msg$ message if the $test\_msg$ message sender is belonging to the same cluster.
 \item $inform\_msg$ (Inform message): cluster $root$ sends this message to the node in which the $moe$ is attached.
 \item $merge\_{req}$ (Merge Request): Merging request from one cluster to some other cluster, containing {\it cluster id}.
 \item $internal\_msg$ (Internal message): This message is for pass the information in the same cluster.
\item $merge\_msg$ (Merge message): To ensure merging between two cluster.
\item  $commit\_msg$ (Commit message): Final commitment
\item $ignore\_msg$ (Ignore message): Ignore requests.
\item $modify\_msg$ (Modify message): This message is generated by the end points of minimum outgoing edge after
merging and pass in the new cluster to find the new $root$.
\end{itemize}

\section{Description of the Algorithm}

Suppose a sensor node with degree $k$ is faulty in the initial aggregation tree $T_{a}$.
Removal of this faulty node decomposes the aggregation tree into $k$ number of trees (or clusters), $T_{1},T_{2}, \cdots, T_{k}$ (say).
Then let us assume by some fault detection algorithm the node, $v_{i}^{df}$ ($root$ of the cluster, say) directly
attached with the faulty node in each cluster, $T_{i}$ can find the information about the fault and starts following reconstruction process.

\subsection{Subround-I: Minimum outgoing edge ($moe$) finding}
For each cluster $T_{i}$, $v_{i}^{df}$ named as $root$ node initiates and sends $find\_msg$ to its descenders within the cluster through the tree edges with the {\it id} of the $root$, named as $T_i^{id}$, which is same as $v_{i}^{df}$.
After receiving $find\_msg$ every other nodes assign $T_i^{id}$ to its local variable ($cluster\_id$) and forwards the message to neighbors
until it reach to leaf nodes. After receiving $find\_msg$ leaf node finds the $moe$ and returns a $report\_msg$ to the sender of $find\_msg$.
After receiving $report\_msg$ all intermediate nodes modify $moe$ if possible with respect to its own $moe$ and forward the $report\_msg$ to the $root$ node. For finding $moe$ a node passes $test\_msg$ with $cluster\_id$ through the possible $moe$ to test whether the other
end of this $moe$ is in the different cluster. If the other end of $moe$ is in different cluster than the node returns a $accept\_msg$ with its own id otherwise the node returns a $reject\_msg$.

After receiving $reject\_msg$ this node again tries to find the next possible $moe$ among its neighbours until it receive
a $accept\_msg$ or there is no possible $moe$ edge for node. In that case the node marks all such rejected edges not to use further for $moe$ selection. 
There may be a possibility of multiple $moe$ at any individual node. In this case the node selects $moe$ with minimum id node among the multiple $accept\_msg$.

After receiving $report\_msg$ the $root$ node finally selects a $moe$ for the cluster and sends $inform\_msg$ to the corresponding node $v^{moe}_i$ (say) attached with the $moe$.

\subsection{Subround-II: Merge message passing}

The node, $v^{moe}_i$ of each cluster, $T_{i}$ sends a $merge\_{req}$ message along their
respective $moe$ to some node of $T_{j}$, say.
The decision after receiving $merge\_{req}$ message as following:
There are two cases:
\begin{enumerate}
\item If $root$, $v^{df}_j$ of $T_{j}$ receives $merge\_{req}$ and if the
$cluster\_id$ of $T_{j}$ is less than the $cluster\_id$ of $T_{i}$ then $v^{df}_j$ returns
 an $ignore\_msg$ to $v^{moe}_i$, otherwise $v^{df}_j$ keep the information in its database.
\item If some other node ($v_{j}$) excluding $v^{df}_j$ of $T_{j}$ receives a $merge\_{req}$ and if $cluster\_id$ of $T_{j}$ is less than the $cluster\_id$ of $T_{i}$ then the node $v_{j}$ returns an $ignore\_msg$
 to $v^{moe}_i$, otherwise $v_{j}$ forwards the message ($internal\_msg$) to the $root$ $v^{df}_j$.
\end{enumerate}

\subsection{Subround-III: Decision after receiving a merge messages}
At the end of the previous Subround-II if $root$ of $T_{j}$ for some $j$ receives one or more than one $merge\_{req}$ messages then
it finds the minimum $cluster\_id$ over all messages and sends a $merge\_msg$ to the minimum id cluster and sends
 $ignore\_msg$ to all others directly or via $v_{j}$ node ($v_{j}$ is considered in the case-2 of Subround-II).
Now, if $root$  of $T_{j}$ for some $j$ does not receive any $merge\_{req}$ or receive but pass a $ignore\_msg$
to sender then the $root$ of $T_{j}$ sends a $merge\_msg$ through the $moe$ (chosen in Subround-II) from $v^{moe}_j$ node.

\subsection{Subround-IV: Merging of clusters}
   In this subround  each cluster $T_{i}$, for $i= 1,2,\cdots,k$ some node $v_i$ (including $root$) receives $merge\_msg$
and/or $ignore\_msg$ from $v_j$ (including $root$) of some other cluster $T_{j}$.
If the message is $ignore\_msg$ then drop the message. Otherwise merge these two clusters
in the following ways:

\begin{enumerate}
 \item If $v_i$ sends a $merge\_msg$ to $v_j$ and if $T^{id}_i$ $ < $ $T^{id}_j$ then
$T_{i}$ sends a $commit\_msg$ to $T_{j}$ and
 $T_{j}$ merge with $T_{i}$ by including the edge in the modified aggregation tree.
After that the vertices attached with the edge initiate $modify\_msg$ over
the new cluster $T_{i}^{'}$ (, say) with the information of $v_{i}^{df}$ for the modification of $root$.
If $v_i$ sends a $merge\_msg$ to $v_j$ and if $T^{id}_i$ $ > $ $T^{id}_j$ then $merge\_msg$ is
drop without merging.
\item If $v_i$ does not send a $merge\_msg$ to $T_{j}$ then $v_i$ sends a $commit\_msg$ and a $modify\_msg$
(as a responds) to cluster $T_{j}$ after receiving $modify\_msg$ from its own cluster. Then $T_{j}$ merge with $T_{i}$ by
 including the edge in the modified aggregation tree and $T_{j}$ expand.
\end{enumerate}

\section{The Algorithm}
\begin{algorithmic}
\STATE $G_c=(V,E) \gets $ Communication graph
\STATE $ T_{a} \gets $ Initial aggregation tree
\STATE $ k \gets $Degree of the faulty node, $v_f$
 \STATE \underline{Subround-I}: (Finding $moe$)
  \FOR{each cluster $T_{i}$ ; $i = 1$ to $k$}
      \STATE  $root$, $v_{i}^{df}$ initiates and sends $<find\_msg, T_i^{id}>$
      \FOR{each node $v_i$}
	  \STATE{$cluster\_{id}_i \gets$ $T_i^{id}$}
      \ENDFOR
   \FOR{each node $v_{i}$ (starts from leaf nodes)}
      \STATE passes $test\_msg$ through its $moe \in E^{'}$ of $G^{'}_c$ to some other node $v_{i'}$
      \IF {$T_i^{id}$ $ \neq $ $T_{i'}^{id}$}
        \STATE $v_{i'}$ returns an $accept\_msg$ to $v_{i}$
	\STATE $v_{i}$ passes a $reprot\_msg$ to the $find\_msg$ sender
      \ELSE
         \STATE $v_{i'}$ returns $reject\_msg$ to $v_{i}$ and marks this rejected edge in $E^{'}$ and
         $v_{i}$ looks for the next possible $moe$
      \ENDIF
    \ENDFOR
\FOR{each node $v_{i}$ (intermediate/$root$)}
\STATE After receiving $report\_msg$ the node modifies $moe$ if possible wrt its own $moe$ as above and forwards $report\_msg$ to its ancestor until it reaches to the $root$
\STATE When $root$ receives the $v^{moe}_i$ then it passes the $inform\_msg$ to the $v^{moe}_i$ if $moe$ is not attached with the $root$
\ENDFOR
\ENDFOR
\IF {there is no $moe$}
\RETURN Tree is reconstructed \& the protocol is terminated
\ELSE
      \STATE moves for the Subround-II
\ENDIF
\end{algorithmic}

\begin{algorithmic}
        \STATE \underline{Subround-II}: (Merge message passing)
         \FOR{each cluster $T_{i}$ ; $i = 1$ to $k$}
	   \STATE $v^{moe}_i$ sends a $merge\_req$ from cluster $T_{i}$  to some $T_{j}$
         \ENDFOR
         \IF  {$v_{j}^{df}$ of $T_{j}$ receives this $merge\_req$ message}
 	    \IF {$C^{id}_{j}$ $ < $ $C^{id}_{i}$}
 		\STATE passes an $ignore\_msg$ to $v^{moe}_i$
 	    \ELSE
 		 \STATE keeps the message
 	    \ENDIF
 	 \ELSE
 	     \IF {some other node $v_{j}$ of $T_{j}$ receives this $merge\_req$ message}
 		  \IF {$C^{id}_{j}$ $ < $ $C^{id}_{i}$}
 		       \STATE passes an $ignore\_msg$ to $v^{moe}_i$
 		  \ELSE
 			\STATE $v_{j}$ receives this message and passes it to $v_{j}^{df}$ of $T_{j}$ through an $internal\_msg$  			
 		  \ENDIF
 	      \ENDIF
          \ENDIF
\end{algorithmic}

\begin{algorithmic}
	  \STATE \underline{Subround-III}: (Decision after receiving a merge messages)
	  \FOR{each cluster $T_{i}$ ; $i = 1$ to $k$}
 	      \IF {$v_{i}^{df}$ receives $merge\_req$ from some other clusters}
		  \STATE sends $merge\_msg$ to the minimum id cluster among them and $ignore\_msg$ to others
	      \ELSE
		   \STATE sends $merge\_msg$ from $v^{moe}_i$ through $moe$
	      \ENDIF
	   \ENDFOR
\end{algorithmic}

\begin{algorithmic}
            \STATE \underline{Subround-IV}: (Merging of clusters)
             \STATE $v^i$ of $T_{i}$ receives either $merge\_msg$ or/and
 		  $ignore\_msg$ from $T_{j}$ after the end of Subround-III
              \IF {message is $ignore\_msg$}
 		\STATE drops the message without merging
  	     \ELSE
  	         \IF {$T_{i}$ also sends a $merge\_msg$ to $T_{j}$}
		     \IF{ $T^{id}_{i}$ $ < $ $T^{id}_{j}$}
			\STATE $T_{i}$ passes a $commit\_msg$ to $T_{j}$
 		        \STATE  $T_{j}$ merges with $T_{i}$ in some new cluster $T^{'}_{i}$
			\STATE the nodes attached with merged edge initiates and sends $modify\_msg$ within $T^{'}_{i}$.
			\STATE $cluster\_id$ of $v_{i'}\in T^{'}_{i}$ resets the value by $v_{i}^{df}$
 		      \ELSE
 		         \STATE drops this received $merge\_msg$
                     \ENDIF
                  \ELSE
		     \STATE $v^i$ sends a $commit\_msg$ and forwards $modify\_msg$ (as a responds) to cluster $T_{j}$ after
                                receiving $modify\_msg$ from its own cluster and then $T_{j} $ merges with $T_{i}$
 		 \ENDIF
 	     \ENDIF
  \STATE 
  Re-execute the protocol from Subround-I with modified clusters until termination.
\end{algorithmic}

\section{Complexity Analysis}

Let $k$ be the number of clusters after a node failure. We are measuring the complexity of the proposed
algorithm in terms of rounds of execution and total number of message exchange.
First we concentrate over possible best and worst rounds of execution.
\begin {itemize}
 \item Case-1 (Best Case) If $v^{moe}_i$ sends $merge\_req$ to the minimum id cluster $T_j$ (,say) for all $i\in \{1,2,\cdots,k\}\setminus \{j\}$, then the tree would be reconstructed in one round.
 \item Case-2 (Worst Case) If every distinct pair of clusters exchange $merge\_req$ in Subround-II and merge in Subround-IV then in one round number of cluster reduces by half. If this kind of merging process is continue then after $O({\rm lg}~ k$) rounds the tree would be reconstructed.
\end {itemize}
Now we determine an upper bound for the number of messages for a cluster $T_i$.

Let the number of nodes in this cluster is $n_i$. Recall the types of messages used by the algorithm :\newline
$find\_msg$: $n_i-1$ $find\_msg$ messages.\newline
 $test\_msg$: (successful test and failed test.)\newline
$accept\_msg$: Acceptance requires two messages, successful test and accept. So the messages are $2n_i$. Note that $inform\_msg$
also included in this count.\newline
 $reject\_msg$:  Note that an edge can be reject at most once throughout the execution
of the algorithm. Rejection requires two messages: failed test and reject. So we have $2E$ messages.\newline
 $report\_msg$: $n_i-1$ $report\_msg$.\newline
$merge\_req$:  1 (one) request for merging.\newline
$ ignore\_msg$: at most $k-1$ $ignore\_msg$ throughout the execution of the algorithm. \newline
 $internal\_msg$: at most $n_i-1$ message.\newline
 $merge\_msg$:  one message.\newline
  $commit\_msg$: one message for final commitment \newline
 $modify\_msg$: $n_i-1$ messages for modification.

 The total number of message required for a cluster is $6n_i$.
 Total number of message for all $k$ clusters is $\sum_{i=1}^{k}(6n_i)$ = $6(n-1)$ where $n-1=\sum_{i=1}^{k}n_i$

 Therefore the total number of message for merging of all $k$ clusters
 is $O(n {\rm lg}~ k+E)$. Here $k$ may be $n-1$, therefore the total counting brings us to $O(n{\rm lg}~ n+E)$.\\

\section{Correctness}
Note that in a single round of proposed algorithm, every cluster sends a unique $merge\_msg$ through $moe$. In
the merging of two or more than two clusters simultaneously there is exactly two clusters
which sends a $merge\_msg$ to each other through the same $moe$.
\begin{theorem} \label{thm2} There is no cycle after merging two or more clusters.
\begin{proof}Let $T_{a}$ be the initial aggregation tree with $n$ nodes and $v_{f}$ be the faulty node. Proof
by induction on degree of $v_{f}$ node in $T_{a}$.\newline
{\it Basis:}
 Let deg($v_{f}$) = $1$. Then after removing $v_{f}$ from $T_{a}$, there is only one
   cluster with $n-1$ nodes. Clearly $T_{a}^{'}$ with $n-1$ nodes is again a tree.\newline
Let deg($v_{f}$) = $2$ and $T_{i}$, $T_{j}$ be the clusters. Let us suppose cycle occurs
in the merging of $T_{i}$ and $T_{j}$. It is possible if both $T_{i}$ and $T_{j}$ send a $merge\_msg$ to each other
through different multiple $moe$. But this contradicts Subround-III of the proposed algorithm. Since according to
proposed algorithm both $T_{i}$ and $T_{j}$ send a $merge\_msg$ to each other through same $moe$. Hence there
is no cycle in the merging of $T_{i}$ and $T_{j}$.\newline
{\it Inductive hypothesis:}
Let no cycle occurs in the merging of $k$ or less clusters, i.e., deg$(v_{f}) \leq k$.\newline
{\it Inductive step:} Now let deg($v_{f}$) = $k+1$ and $T_{i}$, for $i= 1,2,\cdots,k+1$ be the clusters.
Let us suppose cycle occurs in the merging of these $k+1$ clusters. It is possible if at least three cluster
  $T_{1}$, $T_{2}$, $T_{3}$ (, say) send the $merge\_msg$ to each other as $T_{1}$ to $T_{2}$, $T_{2}$ to $T_{3}$, $T_{3}$
 to $T_{1}$ in a round. But this contradicts our algorithm that there are exactly two clusters
which send a $merge\_msg$ to each other through the same $moe$ in the merging of more than two clusters. Therefore cycle
cannot occur in a round and number of clusters reduces. Now by
 inductive hypothesis cycle will not occur in the merging of $k+1$ clusters. Hence theorem is true for
any number of clusters.
\end{proof}
\end{theorem}
\begin{theorem} \label{thm1} Resultant reconstructed aggregation tree is again a MST.
\begin{proof} Let $T_{a}$ be the initial aggregation tree and given that is a MST with $n$
nodes and $v_{f}$ be the faulty node with degree $k$.
Let $T_{a}^{'}$ be the aggregation tree which is reconstructed using our proposed algorithm with $n-1$ nodes after removing the faulty node $v_{f}$ .
Since $T_{a}$ is a MST, therefore removal of $v_{f}$ divides it in to $k$ sub trees where each of them are individually a MST.
Now suppose $T_{a}^{'}$ is not a MST, it means there are at least two clusters which is not merged
with a minimum weighted edge in the $T_{a}^{'}$. But it is a contradiction of our algorithm that
allows merging between different clusters through a minimal weighted edge. Hence the resultant reconstructed aggregation tree
is again a MST.
\end{proof}
\end{theorem}

\section{Multiple Sensor Nodes Failure}

If $m$ number of nodes fail simultaneously and if $d_1, d_2,\cdots, d_m$ are the degrees of respective faulty nodes then
at most $d_1+d_2+\cdots+d_m$ number of disjoint forest may form. Then same proposed algorithm can merge all disjoint forest
and reconstruct the aggregation tree.

\section{Conclusion}
In this paper, we have proposed a distributed algorithm for reconstruction of aggregation tree
in wireless sensor networks when an arbitrary sensor node fails during aggregation. Our model
is synchronous, performing in rounds. In terms of rounds the time complexity of our
algorithm is $O(1)$ in the best case, $O({\rm lg}~ n$) in the worst case.
Our proposed algorithm can also handel multiple concurrent sensor node failure. But the proposed algorithm cannot
handel node failure during the reconstruction phase.
In our future works we will try to incorporate node failure during the reconstruction phase as well.

\bibliographystyle{IEEEtran}
\bibliography{submitted_version_Sep12.bbl}

\end{document}